\documentstyle[preprint,aps]{revtex}
\tighten
\draft
\widetext
\input epsf
\preprint{CLNS 97/1483, HUTP-97/A020, NUB 3159}
\bigskip
\bigskip
\begin{document}
\title{A Chiral $N=1$ Type I Vacuum in Four Dimensions\\
and Its Heterotic Dual}
\medskip
\author{Zurab Kakushadze$^{1,2}$\footnote{E-mail: 
zurab@string.harvard.edu} and 
Gary Shiu$^3$\footnote{E-mail: shiu@mail.lns.cornell.edu}}

\bigskip
\address{$^1$Lyman Laboratory of Physics, Harvard University, Cambridge, 
MA 02138\\
$^2$Department of Physics, Northeastern University, Boston, MA 02115\\
$^3$Newman Laboratory of Nuclear Studies, Cornell University,
Ithaca, NY 14853-5001}
\date{May 21, 1997}
\bigskip
\medskip
\maketitle

\begin{abstract}
{}In this paper we consider Type I string theory compactified on a ${\bf Z}_7$ 
orbifold. The model has $N=1$ supersymmetry, a $U(4) \otimes U(4) \otimes U(4)
\otimes SO(8)$ gauge 
group, and chiral matter. There are only $D9$-branes (for which we discuss 
tadpole cancellation conditions) in this model 
corresponding to a perturbative heterotic description in a certain region 
of the moduli space. We construct the heterotic dual, match the perturbative 
type I and heterotic tree-level massless spectra via giving certain scalars 
appropriate vevs, and point out the crucial role of the perturbative 
superpotential (on the heterotic side) for this matching. The relevant 
couplings in this superpotential turn out to be non-renormalizable 
(unlike the $Z$-orbifold case discussed in Ref \cite{z3}, where Yukawa 
couplings sufficed for duality matching). We also discuss the role of the 
anomalous $U(1)$ gauge symmetry present in both type I and heterotic models. 
In the perturbative regime we match the (tree-level) moduli spaces of these 
models. We point out possible generalizations of the ${\bf Z}_3$ and 
${\bf Z}_7 $ cases to include $D5$-branes which would help in
understanding non-perturbative five-brane dynamics on the heterotic side. 
\end{abstract}

\section{Introduction}

{}In recent years non-perturbative string dynamics has been coming under 
greater control. String dualities have been playing an important role in 
this process, as they allow us to address non-perturbative issues in a given 
string theory by studying them perturbatively in a dual theory.
Supersymmetry has been a key ingredient of string duality, as the larger the 
number of unbroken space-time supersymmetries, the better handle we have 
over non-perturbative string dynamics.
Thus, much progress has been made in understanding $N=4$ and $N=2$ string 
dualities, and now the attention is shifting toward grasping $N=1$ cases. 

{}$N=1$ type I-heterotic duality in four dimensions is a promising arena 
for testing the validity of the idea of $N=1$ string dualities, as well as 
for developing tools that might help understand non-perturbative effects 
in, say, heterotic string theory ({\em e.g.}, dynamics of five-branes 
responsible for enhanced gauge symmetries). The tree-level relation 
between type I and heterotic dilatons in $D$ space-time 
dimensions \cite{Sagnotti} (which follows from the conjectured 
type I-heterotic duality in ten dimensions
\cite{typeI-het-10}) reads:
\begin{equation}
 \phi_H={{6-D}\over 4} \phi_I -{{D-2}\over 16}\log[\det(g_I)]~.
\end{equation} 
Here $g_I$ is the internal metric of the type I compactification space, 
whereas
$\phi_I$ and $\phi_H$ are the type I and heterotic dilatons, respectively. 
One implication of the above equation is that in four dimensions there 
always exists a region in the moduli space where both type I and heterotic 
string theories are weekly coupled, and there we can rely on perturbation 
theory. If we understand the map betwixt perturbative effects in the two 
descriptions, we may be able to learn about non-perturbative effects in, 
say, heterotic string via casting them 
into perturbative effects in type I theory ({\em e.g.}, non-perturbative 
dynamics of heterotic five-branes can presumably be understood by studying 
perturbative dynamics of type I $D5$-branes).

{}Recently one of us studied an example of a four-dimensional $N=1$ 
type I-heterotic dual pair \cite{z3}. The type I model, as well as the 
candidate heterotic dual, considered in Ref \cite{z3} were first constructed 
in Ref \cite{Sagnotti}. The type I model is a compactification on 
the $Z$-orbifold (and has $D9$-branes only), whereas the candidate heterotic 
dual is a $Z$-orbifold compactification with a non-standard embedding of 
the gauge connection. At the orbifold points the tree-level massless 
spectra of the two models differ as there are extra twisted matter fields 
in the heterotic model that do not have (perturbative) type I counterparts. 
As discussed in Ref \cite{z3}, there is a tree-level superpotential in the 
heterotic model precisely such that the extra states become heavy after 
appropriate Higgsing. The role of the anomalous $U(1)$ (present in both 
type I and heterotic models) was also discussed in Ref \cite{z3}.

{}The case studied in Ref \cite{z3} is remarkable in the sense that the 
type I model has only $D9$-branes and the dynamics is completely 
perturbative from the heterotic point of view, hence there is 
not much difficulty 
in establishing (tree-level) duality. In the context of our previous 
discussion, it would be important to see if there is any pattern in such 
perturbative $N=1$ type I-heterotic duality in four dimensions. If so, 
this would help separate perturbative effects from non-perturbative ones 
in the cases with $D5$-branes (which are more involved from the
heterotic point 
of view, and these are the cases we would ultimately like to understand). 
There turns out to be one other case of type I ${\bf Z}_N$ orbifold 
compactification 
with $N=1$ supersymmetry and no $D5$-branes. This is the compactification 
on the ${\bf Z}_7$ orbifold that we study in this paper. 

{}Before discussing the ${\bf Z}_7$ case, we list some orbifolds of 
type I strings with $N=1$ supersymmetry in four dimensions (there are two 
inequivalent ${\bf Z}_6$ orbifolds in 4D):\\
$\bullet$ $D9$-branes only: ({\em i}) ${\bf Z}_3$, ({\em ii}) ${\bf Z}_7$;\\
$\bullet$ $D9$-branes and $D5$-branes: ({\em iii}) ${\bf Z}_6$, ({\em iv}) 
${\bf Z}^\prime_6$,
({\em v}) ${\bf Z}_2 \otimes {\bf Z}_2$, ({\em vi}) ${\bf Z}_4$.\\
So far, only ${\bf Z}_3$ \cite{Sagnotti} and 
${\bf Z}_2 \otimes {\bf Z}_2$ \cite{BL} cases have been constructed, and 
only the ${\bf Z}_3$ case has been studied from the type I-heterotic 
duality point of view. In this paper we discuss the ${\bf Z}_7$ case. The
model has $N=1$ supersymmetry, 
$U(4)\otimes U(4)\otimes U(4)\otimes SO(8)$ gauge group, and chiral matter. 
There is anomalous $U(1)$ in this model. We also construct the heterotic 
dual that has the same gauge symmetry and matter content as the type I 
model except for extra twisted matter fields. This is just as in the 
${\bf Z}_3$ model, albeit there are some subtle differences in the way 
these extra matter fields are charged under the gauge group. 
Another difference 
between the ${\bf Z}_3$ and ${\bf Z}_7$ cases is that the orbifold blow-up 
modes in the former case are charged under the anomalous $U(1)$ and 
contribute to cancelling the $D$-term, whereas in the latter case the 
orbifold blow-up modes are neutral under the anomalous $U(1)$. This results in
 different pictures for embedding of the type I moduli space into that of 
heterotic string in the ${\bf Z}_3$ and ${\bf Z}_7$ cases. Just as in the 
${\bf Z}_3$ case, in the ${\bf Z}_7$ case there is a tree-level 
superpotential that after appropriate Higgsing gives masses to all the 
extra twisted matter in the heterotic model, and the massless spectra of 
the type I and heterotic strings are matched. There is also a difference, 
however: in the ${\bf Z}_3$ case renormalizable (Yukawa) couplings are
sufficient for duality matching, whereas in the ${\bf Z}_7$ case 
the corresponding 
couplings are {\em non-renormalizable}. (In deducing the heterotic 
superpotential the tools developed in Ref \cite{KST} prove to be very 
useful; see Appendix \ref{bosonic} and Appendix \ref{space} for details.) 
With these subtle differences, the 
type I-heterotic duality in the two cases (${\bf Z}_3$ and ${\bf Z}_7$) 
works much in the same way, and there is, henceforth, a clear pattern we 
see from studying these examples. (Note that perturbative superpotentials 
also seem to be necessary for matching the massless spectra of $F$ theory 
and heterotic dual pairs \cite{Bershadsky}.)

{}The paper is organized as follows. In Sec. II we discuss the 
${\bf Z}_7$ orbifold type I model.
In Sec. III we construct the heterotic dual. In Sec. IV we give 
perturbative superpotentials for these models. In Sec. V we discuss 
the moduli space, and explain the matching between the type I and heterotic 
moduli spaces, as well as their tree-level spectra. In Sec. VI we give 
conclusions and remarks. Some of the details regarding the tadpole 
cancellation in type I theory, and also the heterotic superpotential are 
relegated to the Appendices.

\section{Type I Model}

{}In this section we discuss the construction of the type I model.
Let us start from the type IIB string model compactified on the six-torus 
which has a ${\bf Z}_7$ rotational symmetry. (A more 
detailed discussion of this six-torus will be given in the next section 
where we go through the construction of the (candidate) 
heterotic dual of the model considered in this
section.) This model has $N=8$ supersymmetry. Let us now consider the 
symmetric 
${\bf Z}_7$ orbifold model generated by the twist
\begin{equation}
 T_7=(\theta,\theta^2,\theta^3 \vert\vert \theta, \theta^2, \theta^3)~.
\end{equation} 
Here $\theta$ is a $2\pi /7$ rotation of a complex boson (we have complexified 
the six real bosons into three complex bosons). The double vertical 
line separates the right- and left-movers of the string. The resulting model 
has $N=2$ space-time supersymmetry. This model has the following moduli. 
There are 8 Neveu-Schwarz--Neveu-Schwarz (NS-NS) fields 
$\phi,B_{\mu\nu},B_{i{\bar i}},g_{i{\bar i}}$, 
and 8 Ramond-Ramond (R-R) fields 
$\phi^\prime,B^\prime_{\mu\nu},B^\prime_{i{\bar i}},
C^\prime_{\mu\nu i{\bar i}}$.

{}Let us now consider the orientifold projection of this model. The closed 
string sector (which is simply the subspace of the
Hilbert space of the original type IIB spectrum invariant under the orientifold
projection $\Omega$) contains the $N=1$ supergravity multiplet, and 
3 untwisted 
(the NS-NS fields that survive the $\Omega$ projection are 
$g_{i{\bar i}}$, whereas the R-R fields that are kept are 
$B^\prime_{i{\bar i}}$; note that the  
NS-NS field $\phi$ and the R-R field $B^\prime_{\mu\nu}$ also survive and 
enter in the dilaton supermultiplet)
and 21 twisted chiral supermultiplets (which are neutral under the gauge 
group of the model). For consistency (tadpole cancellation; 
see Appendix \ref{tadpoles} for details) we must include 
the open string sector. Note that in this model we only have $D9$-branes 
but no $D5$-branes since the orbifold group does not contain an order 
two element. (If the orbifold group contains an order two element $R$, 
then the sector $R\Omega$ would contain $D5$-branes). Thus, we only have 
$99$ open strings. The gauge group consistent with tadpole cancellation 
then is $U(4)\otimes U(4) \otimes U(4) \otimes SO(8)$. The $99$ open 
strings also 
give rise to the chiral matter fields
$({\bf 4}, {\bf 1}, {\bf 1}, {\bf 8}_v)(+1,0,0)_L$, 
$(\overline{\bf 4}, \overline{\bf 4}, {\bf 1}, {\bf 1})(-1,-1,0)_L$,
$({\overline{\bf 4}}, {\bf 4}, {\bf 1}, {\bf 1})(-1,+1,0)_L$, and
$({\bf 6}, {\bf 1}, {\bf 1}, {\bf 1})(+2,0,0)_L$. In addition, there 
are fields that can be obtained by permuting the three $U(4)$'s [this 
permutation must be accompanied by changing the
irrep of the third $U(4)$ 
to its complex conjugate].
Here the first four entries in bold font indicate the irreps of the
$SU(4)\otimes SU(4) \otimes SU(4) \otimes SO(8)$ subgroup, whereas the 
$U(1)^3$ charges  are given in the parenthesis. 
The subscript $L$ indicates the space-time helicity of the corresponding 
fermionic fields. The massless spectrum of this model is summarized in Table I.

{}Note that the $U(1)^3$ gauge symmetry is anomalous. We can form a linear
combination of these $U(1)$'s such that only one of them
is anomalous [this combination is given by $Q_1+Q_2-Q_3$, where $Q_{1,2,3}$ 
are the first, second, and third $U(1)$ charges, respectively]. The total 
$U(1)$ anomaly is $+36$. 
By the generalized Green-Schwarz
mechanism \cite{GS,DSW} some of the fields charged under $U(1)$ will acquire 
vevs to cancel the Fayet-Illiopoulos $D$-term.

\section{Heterotic String Model}

{}In this section we give the construction of
the heterotic string model that is (candidate) dual to the type I
model considered in the previous section. Let us start from the Narain model
with $N=4$ space-time supersymmetry in four dimensions. Let the momenta of the
internal (6 right-moving and 22 left-moving) world-sheet bosons span the 
(even self-dual) Narain lattice $\Gamma^{6,22}=\Gamma^{6,6}\otimes\Gamma^{16}$.
Here $\Gamma^{16}$ is the ${\mbox{spin}}(32)/{\bf Z}_2$ lattice, whereas the 
lattice $\Gamma^{6,6}$ is spanned by the momenta $(p_R \vert\vert p_L)$ with
\begin{eqnarray}
 p_{L,R}={1\over 2}m_i {\tilde e}^i \pm n^i e_i ~.
\end{eqnarray}
Here $m_i$ and $n^i$ are integers, $e_i \cdot e_j =g_{ij}$ is the constant 
background metric of the compactification manifold (six-torus), and $e_i \cdot
{\tilde e}^j={\delta_i}^j$. Note that we could have included the constant
anti-symmetric background tensor field $B_{ij}$, but for now we will set it 
equal to zero for the reasons that will become clear in the following 
(see Appendix \ref{tadpoles}
for details).

{}This Narain model has the gauge group $SO(32) \otimes U(1)^6$. The first 
factor $SO(32)$ comes from the $\Gamma^{16}$ lattice (the $480$ roots of 
length squared 2), and 16 oscillator excitations of the corresponding 
world-sheet bosons [the latter being in the Cartan subalgebra of $SO(32)$].
The factor $U(1)^6$ comes from the oscillator excitations of the six 
left-moving world-sheet bosons corresponding to $\Gamma^{6,6}$. Note that
there are also six additional vector bosons coming from the oscillator 
excitations of the right-moving world-sheet bosons corresponding to 
$\Gamma^{6,6}$. These vector bosons are part of the $N=4$ supergravity 
multiplet.

{}Next consider the ${\bf Z}_7$ orbifold model (with non-standard embedding
of the gauge connection) obtained via twisting the above Narain model by the
following ${\bf Z}_7$ twist:
\begin{equation}
 T_7=(\theta,\theta^2,\theta^3 \vert\vert \theta,\theta^2,\theta^3 \vert
 (\textstyle{1\over 7})^{4} (\textstyle{2\over 7})^4 
 (\textstyle{3\over 7})^4 0^4)~.
\end{equation}   
Here $\theta$ is a $2\pi /7$ rotation of a complex boson (we have complexified 
the original six real bosons into three complex ones). Thus, the first three
entries correspond to the ${\bf Z}_7$ twists of the three right-moving
complex bosons (coming from the six-torus).
The double vertical line separates the right- and left-movers.
The first three left-moving entries correspond to the ${\bf Z}_7$ twists of 
the three left-moving complex bosons (coming from the six-torus). The single 
vertical line separates the latter from the sixteen real bosons corresponding 
to the $\Gamma^{16}$ lattice. The latter are written in the $SO(32)$ basis. 
Thus, for example, $(+1, -1, 0^{14})$ is a root of $SO(32)$ with length 
squared $2$. There are $480$ roots similar to this 
in the $\Gamma^{16}$ lattice, 
and they are descendents of the identity irrep of $SO(32)$. The lattice 
$\Gamma^{16}$ also contains one of the spinor irreps as well. Thus, we will 
choose this spinor irrep to contain the momentum states of the form 
$(\pm {1\over2},...,\pm {1\over 2})$ with even number of plus signs.

{}Note that for the above twist to be a symmetry of the model it is necessary 
(and sufficient) that the twist acting on the $\Gamma^{6,6}$ lattice is a 
rotation in this lattice. This requirement constrains the possible values of 
the metric tensor $g_{ij}$. 

{}Now we are ready to discuss the orbifold model generated by the above twist
$T_7$. This model has $N=1$ space-time supersymmetry, and gauge group
$U(4)\otimes U(4) \otimes U(4) \otimes SO(8)$, the same as the type I model 
discussed in the previous 
section. The untwisted sector gives rise to the $N=1$ supergravity multiplet
coupled to the $N=1$ Yang-Mills gauge multiplet in the adjoint of 
$U(4)\otimes U(4) \otimes U(4) \otimes SO(8)$.
The matter fields in the untwisted sector are the 
same as those in the open string sector of the type I model.
There are also chiral multiplets
neutral under the gauge group: $3({\bf 1},{\bf 1},{\bf 1},{\bf 1})(0,0,0)_L$. 
Note that these contain
six scalar fields that are the left-over geometric moduli whose vevs 
parametrize 
the moduli space $[SU(1,1,{\bf Z})\backslash SU(1,1)/ U(1)]^3$. 
[This is the subspace
of the original Narain moduli space 
$SO(6,6,{\bf Z})\backslash SO(6,6)/SO(6)\otimes 
SO(6)$ that is invariant under the twist.] Actually, the (perturbative) 
moduli space of this model is larger, and we will return to this point 
later on.

{}Next, consider the twisted sector. In the twisted sector we have the 
chiral supermultiplets 
$7({\bf 1}, {\bf 1}, {\bf 1}, {\bf 1})(4/7, 8/7, -12/7)_L$ and
$7({\bf 1}, {\bf 1}, {\bf 6}, {\bf 1})(4/7, 8/7, 2/7)_L$ together 
with fields obtained by permuting the three $U(4)$'s [this permutation must 
be accompanied by changing the irrep of the third $U(4)$ to its complex 
conjugate].
Here we note that the factor 7 comes from the number of fixed points
of the ${\bf Z}_7$ orbifold we are considering. 

{}We summarize the massless spectrum of this heterotic string model 
in Table II.
Note that the $U(1)^3$ gauge symmetry is anomalous. Again, only one linear
combination of the three $U(1)$'s is anomalous. Thus, the contributions of the 
untwisted and twisted sectors into the trace anomaly are
$+36$ and $7 \times (+36)$, respectively,
so that the total trace anomaly is $+288$. By the generalized Green-Schwarz 
mechanism \cite{GS,DSW} some of the fields charged under $U(1)$ will acquire 
vevs to cancel the Fayet-Illiopoulos $D$-term. 

\section{Superpotential}

{}In this section we discuss the perturbative superpotentials for the type I 
and heterotic string models discussed in the previous sections. Studying the 
couplings and flat directions in these superpotentials will enable us to 
make the type I-heterotic duality map more precise.

{}Let us start from the type I model of Sec. II.
We refer the reader to Table I for the massless spectum as well as
our notation. 
Note that perturbatively the 
24 chiral singlets coming from the closed string sector are flat. 
This can be explicitly seen by computing the scattering amplitudes for 
these modes within the framework of the conformal field theory of 
orbifolds \cite{DFMS}. On the 
other hand, the matter fields coming from the $99$ open string sector 
have three (and, of course, some higher) point couplings. The lowest order 
superpotential can be written as (the calculation of the type I 
superpotential is completely analogous to that of the heterotic one in the 
untwisted sector)
\begin{equation}
 W_I =  \lambda_{1} \epsilon_{abc} {\mbox{Tr}} (P_a P_b Q_c) 
      + \lambda_{2} {\mbox{Tr}} 
       (Q_1 R_2 \Phi_3 + Q_2 R_3 \Phi_1 + Q_3 R_1 \Phi_2) 
      + \lambda_{3} {\mbox{Tr}} (R_1 R_2 R_3) + \cdots ~.
\end{equation}
Due to the presence of the anomalous $U(1)$, some of the fields that are 
charged under this $U(1)$ (namely, $Q_a$) 
must acquire vevs to cancel the Fayet-Illiopoulos $D$-term. 
This results in breakdown of gauge symmetry, yet the space-time 
supersymmetry is preserved.

{}Now let us turn to the heterotic string model. The superpotential of this 
model is more involved than that of the type I model as there are non-trivial
couplings between the untwisted and the twisted sector fields. 
We refer the reader to Appendix \ref{bosonic} and Appendix \ref{space} for 
the details of calculating these couplings. The superpotential for the 
heterotic string model thus reads (here we are only interested in the 
general structure of the {\em non-vanishing} terms):
\begin{eqnarray}
 W_H =&&\lambda_{1}^{\prime} \epsilon_{abc} {\mbox{Tr}} (P_a P_b Q_c) 
      + \lambda_{2}^{\prime} {\mbox{Tr}} 
       (Q_1 R_2 \Phi_3 + Q_2 R_3 \Phi_1 + Q_3 R_1 \Phi_2) 
      + \lambda_{3}^{\prime} {\mbox{Tr}} (R_1 R_2 R_3)+  
        \nonumber \\
      && \Lambda^{\alpha \beta \gamma}
           {\mbox{Tr}} ( (Q_{1})^2 S^3_{\alpha} T^1_{\beta} T^2_{\gamma}
                       + (Q_{3})^2 S^1_{\alpha} T^2_{\beta} T^3_{\gamma}
                       + (Q_{2})^2 S^2_{\alpha} T^3_{\beta} T^1_{\gamma})
         + \cdots ~.
\end{eqnarray}
(The notation for the fields are given in Table II.) The couplings
$\Lambda^{\alpha \beta \gamma}$ 
are non-vanishing if the orbifold {\em {space group}} selection rules are
satisifed. Here we note that the couplings 
$\Lambda^{\alpha \beta \gamma}$ 
for $\alpha$, $\beta$, $\gamma$ not all identical
are exponentially suppressed in the limit of large volume 
compactification, whereas the couplings 
$\Lambda^{\alpha \alpha \alpha}$  are not suppressed.
This is because in the former case, the corresponding fields are coming from  
different fixed points so that upon taking them apart 
(in the limit of large volume of the orbifold) their coupling becomes 
weaker and weaker.

{}Following the discussion in Appendix \ref{space} we observe that upon 
the fields 
$Q_a$ [that are responsible for breaking of the anomalous $U(1)$] and 
$S^a_{\alpha}$ (that are the 21 blow-up modes of the ${\bf Z}_7$ orbifold) 
acquiring vev, the states $T^a_\alpha$ generically become heavy and decouple 
from the massless spectrum. Thus, after blowing up the orbifold singularities 
on the heterotic side combined with some of the untwisted charged matter 
fields acquiring vevs to cancel the $D$-term, we can match the massless 
spectrum to that of the type I model [where the charged matter must acquire 
vevs to cancel the effect of the anomalous $U(1)$]. Note the crucial role 
of the perturbative superpotential in this matching. It is precisely such 
that all the extra fields on the heterotic side can be made massive. Here 
we note that the blow-up modes $S^a_{\alpha}$ are neutral under the anomalous 
$U(1)$, and thus do not play important role in cancelling the $D$-term. 
(This is to be contrasted with the $Z$-orbifold model discussed in 
Ref \cite{z3}, where the blow-up modes of the $Z$-orbifold 
carried anomalous $U(1)$ charge.)

\section{Moduli Space}

{}We now turn to the discussion of the moduli spaces for the type I and 
heterotic models considered in the previous sections. Let us start with 
the heterotic model. The (perturbative) moduli space of the corresponding 
Narain model before orbifolding is 
$SO(6,22,{\bf Z}) \backslash SO(6,22) /SO(6)\otimes SO(22)$. After 
orbifolding we have two types of moduli: those coming from the untwisted 
sector, and those coming from the twisted sector. The untwisted sector 
moduli parametrize the coset 
$[SU(1,3,{\bf Z}) \backslash SU(1,3)/SU(3) \otimes U(1)]^3$. 
The subspace $[SU(1,1,{\bf Z}) \backslash SU(1,1)/ U(1)]^3$ of this 
moduli space is parametrized by six neutral singlets $\phi_{a}$ that 
correspond to the 
left-over geometric moduli (coming from the constant metric $g_{ij}$ and 
antisymmetric tensor $B_{ij}$ fields). The other $12$ moduli correspond to 
the flat directions in the superpotential for the fields $P_a$, $Q_a$, $R_a$
and $\Phi_a$. 
(These are the left-over moduli coming from the $6\times 16$ Wilson lines 
$A^I_i$, $I=1,...,16$, in the original Narain model).

{}Next, we turn to the twisted moduli of the heterotic string model. 
In the twisted sectors, we have the chiral superfields 
$S_{\alpha}^{a}$ and $T_{\alpha}^{a}$. There is no superpotential for
the singlets $S_{\alpha}^{a}$ which are the $21$ blow-up modes of the 
${\bf Z}_7$ orbifold.
Unlike the ${\bf Z}_3$ case \cite{z3},
the blow-up modes are not charged under the anomalous $U(1)$ and so all
of them survive the Higgsing process. 
Notice that both the heterotic and the type I model have anomalous $U(1)$
with positive trace anomaly. 
To cancel the $D$-term, one needs to give vevs to the corresponding 
negatively charged fields, namely, $Q_a$. 
At a generic point on the heterotic side ({\em i.e.}, 
upon giving appropriate vevs to the untwisted matter fields $Q_a$ and 
the twisted moduli $S_{\alpha}^{a}$), 
the fields $T_{\alpha}^{a}$ become massive (according to the couplings in 
the superpotential).
Thus, the matching is complete after giving 
appropriate vevs to both untwisted and twisted fields on the 
heterotic side, as well as giving appropriate vevs to open string sector 
matter fields, and $21$ twisted closed string moduli. Upon
breaking the anomalous $U(1)$, the dilaton may mix with other gauge singlets.
{\em A priori}, the mixing is different on the type I and the heterotic
side. To 
make the matching precise, one generically has to appropriately tune the 
dilaton plus $\phi_{a}$ geometric moduli on both sides.

{}Let us analyze more carefully how this matching can be achieved.
Upon giving vev to $Q_1$, the second and the third $U(4)$ are broken to the
diagonal $U(4)$.
Some of the fields
become heavy, whereas $Q_1$ is eaten by the super-Higgs mechanism.
The gauge group is further broken to 
$SU(4)_{diagonal} \otimes SO(8) \otimes U(1)$
once $Q_2$ acquires a non-zero vev. To break the anomalous $U(1)$,
generically, the field 
$Q_3=({\bf 6},{\bf 1})(-2) \oplus (\overline{\bf 10},{\bf 1})(-2)$ 
[in the representations of $SU(4)_{diagonal} \otimes SO(8) \otimes U(1)$]
acquires a vev. The final gauge group is $Sp(4)$ or $SO(4)$
depending on whether $({\bf 6},{\bf 1})(-2)$ or 
$(\overline{\bf 10},{\bf 1})(-2)$ acquires a vev.

{}Thus, the moduli spaces (at generic points) of both type I and heterotic 
models are the same (at least at the tree-level). They are described by the 
untwisted moduli of the heterotic string, or equivalently, the moduli 
coming from the untwisted closed string sector and the open string sector 
of the type I model (these parametrize the coset 
$[SU(1,3,{\bf Z}) \backslash SU(1,3)/SU(3) \otimes U(1)]^3$), plus the 
$2\times 21$ twisted moduli in the heterotic string model, or equivalently, 
the moduli coming from the twisted closed string sector of the type I model. 
The (perturbative) moduli space (of the heterotic model) is schematically 
depicted in Fig.1.

{}It is worth noting the role of anomalous $U(1)$ in $N=1$ type I-heterotic
duality. To cancel the Fayet-Illiopoulos $D$-term, fields that are charged 
under the anomalous $U(1)$ will generically acquire vevs. As a result,
the extra twisted matter fields
in the heterotic model are higgsed away and the matching of the
massless spectra of the type I and heterotic models is precise.
The appearance of massless twisted matter fields $T_{\alpha}^{a}$ on 
the heterotic side is a perturbative effect. On the type I side this 
effect is non-perturbative, and reflects the fact that from type I point 
of view there is a (non-perturbative) singularity 
in the moduli space (or, more precisely, 
a singular subspace of the full moduli space).
Notice that the fields $T_{\alpha}^{a}$ in the heterotic model get heavy 
via {\em non-renormalizable} terms in the {\em perturbative} 
superpotential. This indicates the importance
of perturbative superpotential in $N=1$ type I-heterotic duality.

\section{Conclusions}

{}In this paper we discussed a chiral $N=1$ type I model in four dimensions 
obtained as a compactification on the ${\bf Z}_7$ orbifold. We studied the 
type I -heterotic duality in this example, and have concluded that 
(up to model-dependent differences) the duality in the presence of 
$D9$-branes only has a clear pattern to it. Note that ${\bf Z}_3$ 
(studied in Ref \cite{z3}) and ${\bf Z}_7$ (studied in this paper) cases 
exhaust 4D ${\bf Z}_N$ orbifolds of type I strings with 
$N=1$ supersymmetry and no 
$D5$-branes. The cases with $D5$-branes (constructed via orbifolds of 
even order) are of great interest as they may shed light on 
{\em non-perturbative} dynamics of heterotic five-branes in 4D $N=1$ vacua. 
Having learned the perturbative part of type I-heterotic duality, now we can 
march into the more intricate maze of $D9$- and $D5$-branes and their 
(non-)perturbative heterotic duals.

\acknowledgements

{}We would like to thank Michael Bershadsky, Peter Cho, Andrei Johansen, 
Joe Polchinski, John Schwarz, Matthew Strassler, Tom Taylor, Henry Tye and 
Cumrun Vafa 
for discussions.
The research of G.S. was partially supported by the 
National Science Foundation. G.S. would also like to thank
Joyce M. Kuok Foundation for financial support.
The work of Z.K. was supported in part by the grant NSF PHY-96-02074, 
and the DOE 1994 OJI award. 
Z.K. would also like to thank Albert and Ribena Yu for 
financial support.

\appendix
\section{Tadpoles for Orbifold Singularities}\label{tadpoles}

{}In this appendix we discuss the tadpole cancellation constraints for 
orbifold compactifications of type I strings. We confine our attention to 
${\bf Z}_N$ orbifolds without $D5$-branes. This means that $N$ is odd, and 
without loss of generality we can take $N$ to be a prime number (as all 
the ${\bf Z}_N$ orbifold cases with $N=1$ supersymmetry and no $D5$-branes 
in six and four 
dimensions are restricted to prime $N$). The constraints that we present 
here can be easily generalized to other cases (including those with 
$D5$-branes), which will be discussed elsewhere.

{}There are two kinds of constraints we need to consider. The first one 
comes from the cancellation of the untwisted tadpoles for the $D9$-branes. 
This constraint is the same in all dimensions and leads to the statement 
that there are $16$ $D9$-branes {\em not} counting the orientifold images. 
(This last statement is only correct if the NS-NS antisymmetric background 
$B_{ij}$ is set equal to zero; see below.) The other constraint comes from 
the cancellation of the twisted tadpoles for the $D9$-branes. The twisted 
tadpoles have been computed in six dimensions in Ref \cite{GP} (for the 
${\bf Z}_2$ orbifold limit of $K3$) and Ref \cite{GJ} (for all the other 
orbifold limits of $K3$), and in four dimensions for the 
${\bf Z}_2 \otimes {\bf Z}_2$ orbifold \cite{BL}, and the ${\bf Z}_3$ 
orbifold \cite{Sagnotti}. For the case we are considering here (odd 
prime $N$), there is a simple formula which expresses the twisted tadpole 
cancellation 
condition that we are now going to discuss.

{}Let us be general here and consider compactification on $T^{2d}/{\bf Z}_N$ 
with the twist given by
\begin{equation}
 T_N=(t_1,t_2,...,t_d \vert\vert t_1,t_2,...,t_d)~.
\end{equation}
Here  $t_i$ are fractional numbers taking values in 
$\{0,1/N,2/N,...,(N-1)/N\}$. A given $t_i$ corresponds to a twist of the 
$i$-th complex boson by a $2\pi t_i$ rotation. (We have complexified the 
$2d$ real bosons into $d$ complex bosons). The double vertical line separates 
the right- and left-movers of the string. 
Because we are considering symmetric orbifold, the right- and 
left-moving twists are the same. 
Also note that the consistency of the orbifold requires that the expression
\begin{equation}
   \prod_{i=1}^{d} 4\sin^2(\pi t_i)~,
\end{equation}
where the factors with $t_i=0$ are not included in the product, be an 
integer. In fact the latter is nothing but the number of fixed 
points (tori) in the 
$T_N$ twisted sector.

{}The orbifold action on Chan-Paton factors is described by the unitary 
matrices $\gamma$ that act on the string end-points. In our case $\gamma$ 
(we are suppressing all the indices as they are straightforward to 
reconstruct) is a $16\times 16$ matrix (note that it is {\em not} a 
$32\times 32$ matrix because we have chosen {\em not} to count the 
orientifold images of the $D9$-branes). We can diagonalize this matrix. 
Then, the most general form of this matrix is given by
\begin{equation}
 \gamma=\bigotimes_{k=0}^{N-1} \omega^k {\bf I}_{m_k}~.
\end{equation}
Here $\omega\equiv\exp(2\pi i /N)$, and ${\bf I}_{m_k}$ is the 
$m_k \times m_k$ unit matrix. Note that $\sum_{k=0}^{N-1} m_k=16$.

{}The twisted tadpole cancellation condition in our notation reads:
\begin{equation}
 {\mbox{Tr}}(\gamma)=16 p~,~~~p\equiv 
      \prod_{i=1}^{d} (-1)^{Nt_i} \cos(\pi t_i)~.
\end{equation}
Note that  for this equation to have a solution, we must put 
$m_1=m_2=...=m_{N-1}\equiv m$. [Also note that $m_0=16-m(N-1)\equiv n$.] 
Then ${\mbox{Tr}}(\gamma)=16-Nm$ (note that $\sum_{k=1}^{N-1} \omega^k =-1$).

{}The gauge group of the model can be easily seen from the above tadpole 
equation. It is given by $U(2m)^{(N-1)/2} \otimes SO(2n)$, where 
\begin{equation}
 m=16(1-p)/N,~~~n=16[1+(N-1)p]/N~.
\end{equation}
Note that if none of the $t_i$ are zero, then $\vert p \vert=2^{-d}$.

{}Let us illustrate these equations with a few examples:\\
$\bullet$ 6D ${\bf Z}_3$ orbifold ({\em i.e.}, ${\bf Z}_3$ orbifold limit of 
$K3$). The twist reads: 
\begin{equation}
 T_3=(1/3,1/3 \vert\vert 1/3,1/3)~.
\end{equation}
Note that $p=+1/4$ in this case, $m=4$ and $n=8$, so that the gauge group 
is $U(8) \otimes SO(16)$. This is the model considered in Ref \cite{GJ}.\\
$\bullet$ 4D ${\bf Z}_3$ orbifold ({\em i.e.}, $Z$-orbifold limit of 
a Calabi-Yau three-fold). The twist reads: 
\begin{equation}
 T_3=(1/3,1/3,1/3 \vert\vert 1/3,1/3,1/3)~.
\end{equation}
Note that $p=-1/8$ in this case, $m=6$ and $n=4$, so that the gauge 
group is $U(12) \otimes SO(8)$. This is the model considered in 
Ref \cite{Sagnotti}.\\
$\bullet$ 4D ${\bf Z}_7$ orbifold ({\em i.e.}, ${\bf Z}_7$ orbifold limit 
of a Calabi-Yau three-fold). The twist reads:
\begin{equation}
 T_7=(1/7,2/7,3/7 \vert\vert 1/7,2/7,3/7)~.
\end{equation}
Note that $p=+1/8$ in this case, $m=2$ and $n=4$, so that the gauge group 
is $U(4)\otimes U(4) \otimes U(4) \otimes SO(8)$. This is the model 
considered in this paper.\\
Here we also give two non-supersymmetric models (that have never been 
discussed previously to the best of our knowledge):\\
$\bullet$ 6D ${\bf Z}_5$ orbifold ({\em i.e.}, compactification
 on $T^4/{\bf Z}_5$). The twist reads: 
\begin{equation}
 T_5=(1/5,2/5 \vert\vert 1/5,2/5)~.
\end{equation}
Note that $p=-1/4$ in this case, $m=4$ and $n=0$, so that the gauge group 
is $U(8) \otimes U(8)$.\\
$\bullet$ 8D ${\bf Z}_3$ orbifold ({\em i.e.}, compactification 
on $T^2/{\bf Z}_3$). The twist reads: 
\begin{equation}
 T_3=(1/3\vert\vert 1/3)~.
\end{equation}
Note that $p=-1/2$ in this case, $m=8$ and $n=0$, so that the gauge group 
is $U(16)$.

{}Finally, we would like to consider the cases with non-zero NS-NS 
antisymmetric background
$B_{ij}$. Although there are no massless scalars corresponding to these 
in type I theory (recall that there $B_{ij}$ fields are projected out of 
the spectrum after orientifolding), {\em i.e.}, these moduli cannot be 
varied continuously, they can have certain quantized values (because of 
this they are not moduli in the conventional sense of this word). The 
quantization is due to the fact that to have a consistent orientifold the 
corresponding type IIB spectrum must be left-right symmetric. At generic 
values of $B_{ij}$ this symmetry is destroyed. There are, however, certain 
discrete $B_{ij}$ backgrounds compatible with the orientifold 
projection \cite{Sagnotti1}. The effect of non-zero $B_{ij}$ background 
is that the rank of the gauge group coming from the $SO(32)$ ({\em i.e.}, 
Chan-Paton) factor is reduced, depending on the rank $r$ (which is always 
even) of the matrix $B_{ij}$. That is, the number of the $D9$-branes 
required by the tadpole cancellation condition is no longer $16$ but 
$16/2^{r/2}$. All of 
the above formulas then get modified in the presence of rank $r$ $B_{ij}$ 
in an obvious way via replacing the factor $16$ everywhere by $16/2^{r/2}$.

{}As mentioned earlier, it is not difficult to generalize the tadpole 
cancellation condition discussed in this appendix to the cases with 
$D5$-branes. The work on these cases is in progress and will be reported 
elsewhere. (It would also be interesting to generalize the above conditions 
along the lines of Ref \cite{BI}.)

\section{Bosonic Supercurrent and Scattering}\label{bosonic}

{}In this appendix, we review the bosonic supercurrent approach in
calculating scattering amplitudes of orbifold models \cite{KST}.
The basic idea of this approach is that at the enhanced symmetry point, 
we can rewrite the twists as shifts in the momentum lattice. 
The twist fields can be expressed in terms of ordinary momentum states; 
their quantum numbers
are straightforward to identify and their correlation functions are 
easy to calculate. 

{}To be specific, let us focus on four-dimensional heterotic string 
models within the framework of conformal field theory and consider only
Abelian orbifolds. Before orbifolding, the corresponding Narain model has 
$N=4$ space-time supersymmetry and the internal momenta span an even 
self-dual Lorentzian lattice $\Gamma^{6,22}=\Gamma^{6,6} \oplus \Gamma^{16}$. 
Let $X ({\overline z})$ be one of the three right-moving complex bosons 
corresponding to the six compactified dimensions in $\Gamma^{6,22}$. 
In terms of two real bosons, $X=(X_1 +iX_2)/{\sqrt 2}$. For a ${\bf Z}_N$
twist (for simplicity, $N$ is taken to be prime), 
in the neighborhood of a twist field located at the origin, 
$X ({\overline z})$ undergoes a phase rotation 
\begin{eqnarray}\label{monod}
 \partial X ({\overline z}e^{-2\pi i})=\exp(-2\pi i k/N) \partial X
({\overline z})~,
\end{eqnarray}
which is called the monodromy of $X ({\overline z})$. (Note that $k$ is an 
integer.)
The basic twist field $\sigma ({\overline z})$ has conformal weight 
$h=k(1-k/N)/{2N}$. It 
twists $X({\overline z})$ by $\exp(-2\pi i k/N)$ and its 
complex conjugate ${\overline X}({\overline z})$ by  
$\exp(2\pi i k/N)$, ${\em i.e.}$, 
their operator product expansions (OPEs) are \cite{DFMS}
\begin{eqnarray}\label{tau}
 i \partial X ({\overline z}) \sigma ({\overline w}) &=& ({\overline z} -
{\overline w})^{-(1-k/N)} \tau({\overline w}) + \cdots  ~,\nonumber \\
 i \partial {\overline X} ({\overline z}) \sigma ({\overline w}) &=&
({\overline z} -{\overline w})^{-k/N} \tau ' ({\overline w}) + \cdots ~,
\end{eqnarray}
where $\tau $ and $\tau '$ are excited twist fields. 

{}In this paper, we consider compactification on a ${\bf Z}_7$ orbifold.
The lattice $\Gamma^{6,6}$ must have a ${\bf Z}_7$ symmetry. At generic 
points with this symmetry the gauge symmetry (of the Narain model) coming 
from $\Gamma^{6,6}$ is $U(1)^6$. This symmetry is enhanced to $SU(7)$ at 
the special point.
In terms of six real bosons $\phi^{I}$, the Cartan generators are 
$i \partial \phi^{I}$, whereas the root generators are 
$J_{Q}= \exp (i{Q} \cdot {\phi})~ c({Q})$. 
Here, we have introduced six-dimensional real vectors 
${Q}=({Q}^1,...,{Q}^6)$ which are
root vectors of $SU(7)$ with length squared 2. 
The $c({Q})$ are cocycle operators necessary in the 
Kac-Moody algebra. For convenience, we shall not always explicitly display
$c({Q})$: their presence is understood.  

{}Suppose we can rewrite each $i \partial X^{a}$ (where $X^{a}$ are the 
three right-moving complex bosons corresponding to $\Gamma^{6,6}$) as a sum of 
the root generators,
\begin{equation}\label{X^a}
 i\partial X^a =\sum_{{Q}^{2}=2} {\xi}^a ({Q}) 
J_{Q}~,~~~a=1,2,3~.
\end{equation}    
Then a twist on $\partial X^{a}$ in 
Eq. (\ref{monod}) becomes a shift in $\phi^{I}$:
\begin{equation}
 	\phi ^{I}({\overline z}e^{-2\pi i})= \phi^{I} ({\overline z}) 
                  - 2\pi U^{I} ~,
\end{equation}
where $Q \cdot U=k/N$.
The coefficients ${\xi}^a ({Q})$ must be chosen such that
the following OPEs are satisfied:
\begin{eqnarray}\label{XOPE}
\partial X^{a} (\overline{z}) \partial X^{b} (0) &\sim& {\mbox{regular}} 
\nonumber ~,\\
\partial X^{a} (\overline{z}) \partial X^{b \dagger} (0) &\sim& 
- \overline{z}^{-2} \delta^{ab} + {\mbox{regular}} ~.
\end{eqnarray} 
For a lattice with $SU(N)$ enhanced symmetry ($N=7$ in our case), 
the choice is unique (up to equivalent representations),
\begin{equation}\label{SU(N)}
 i \partial X = {1 \over {\sqrt N}} \sum_{\alpha=1}^N 
e^{-ie_\alpha \cdot \phi}~,~~~
 i \partial {\overline X} = {1 \over {\sqrt N}} \sum_{\alpha=1}^N e^{ie_\alpha 
\cdot \phi}~.
\end{equation}
where $e_{\alpha}$ for $\alpha=1,\dots N-1$ are simple roots of $SU(N)$
and $e_{N}=- \sum_{\alpha=1}^{N-1} e_{\alpha}$. We have suppressed the
index $a$. The set of roots chosen depends on the
monodromy of $i \partial X^a$ and can be different for $a=1,2,3$.

{}To see explicitly how this conversion of twists to shifts can
be realized, let us consider the decomposition of 
${\bf N}$ of $SU(N)$ into representations of $U(1)^{N-1}$:
\begin{eqnarray}\label{fund}
{\bf N} =&& (1,\dots,1) \oplus (-1,1,\dots,1) \oplus (0,-2,1,\dots,1)
                   \oplus (0,0,-3,1,\dots,1) \nonumber \\
         &\oplus&  \dots \dots \dots \oplus (0,0,\dots,-N+1) ~.
\end{eqnarray}
The normalization radii of $U(1)^{N-1}$ are given by 
$({1\over \sqrt{1 \cdot 2}},{1\over \sqrt{2 \cdot 3}},\dots,
{1\over \sqrt{(N-1) \cdot  N}})$.
They are chosen such that
all the states on the right-handed side
of the equation have the same conformal dimension as ${\bf N}$ of $SU(N)$,
{\em i.e.}, $h=(N-1)/2N$. 
The conjugate representation ${\overline{\bf N}}$ has opposite $U(1)^{N-1}$
charges. The adjoint (and hence the roots $Q$) of $SU(N)$
can be obtained from the tensor product ${\bf N} \otimes {\overline{\bf N}}$.

{}To construct the shift representation $U$ of the twisted fields in the
singly twisted sector,
we demand that the $N$ states in Eq.~(\ref{fund}) pick up different phases
under the shift ({\em i.e.}, $Q \cdot U = k/N$ for $k=0,1,\dots,N-1$).
The shift representation is uniquely determined once we fix the assignment
of the phases. The roots of $SU(N)$ can be divided into sets with 
different phases under the shift $U$. Terms that appear in Eq.~(\ref{SU(N)})
are determined by the monodromy of $i \partial X$. [Thus, for $N=7$ we 
have $42$ roots. $21=3\times 7$ enter in the expressions (\ref{SU(N)}) 
for the three bosons $i \partial X^a$, and the other $21=3\times 7$ roots 
enter in the expressions (\ref{SU(N)}) for the three bosons (conjugate) 
$i \partial {\overline X}^a$.]

{}For prime $N$, there are $N$ twisted sectors, 
each with $N$ fixed points. To obtain the singly twisted fields corresponding
to the other fixed points, we simply add weights of $SU(N)$ to $U$
such that the conformal dimension is preserved, {\em i.e.}, 
${1\over 2}(U+W)^2 = {1\over 2} U^2$. 
There is precisely one weight vector in each unitary representation 
of $SU(N)$ that satisfies the above requirement, and there are
$N$ unitary representations
of $SU(N)$: the identity ${\bf 1}$ and the antisymmetric 
tensors 
$\epsilon^{ij \cdots k} {\bf N}_i \otimes 
{\bf N}_j \otimes \cdot\cdot\cdot\ \otimes {\bf N}_k$.
[In the case of $SU(7)$, they are the identity ${\bf 1}$, the weights 
${\bf 7}$,
${\bf 21}$, ${\bf 35}$, and their conjugate weight representations.]
The twisted fields in higher twisted sectors are represented by the shifts
$kU + Q$ where $k=2,\dots,N-1$ and $Q$ is a root vector added 
so as to preserve the conformal dimension of the twisted states.
Again, the higher twisted fields at other fixed points are obtained
by adding appropriate weights.

{}In the standard orbifold formalism, the internal part of the 
supercurrent for the right-movers
can be written as
\begin{equation}
 T_F={i\over 2}\sum_{a=1}^{3} \psi^a \partial X^a +{\mbox {H.c.}}~,
\end{equation}
where $\psi^a$ are complex world-sheet fermions.
The twists on $\psi^{a}$ can be written as shifts if we bosonize
the complex fermions:
\begin{eqnarray}
\psi^{a}&=& \exp(i \rho^{a}) =\exp(i H \cdot \rho) ~, \nonumber \\
\psi^{a \dagger}&=& \exp(-i \rho^{a}) =\exp(-i H \cdot \rho) ~.
\end{eqnarray}
where $H$ (known as the $H$-charge) equals $(1,0,0)$, $(0,1,0)$ or $(0,0,1)$ 
for $a=1,2,3$.

{}The bosonic supercurrent is then given by (the cocycle operators are not
displayed)
\begin{equation}
T_F = {1\over 2 \sqrt{N}} \sum_{a=1}^{3} e^{i \rho^{a}} 
                          \sum_{\alpha=1}^{N} e^{i Q^a_{\alpha} \cdot \phi}
                          +{\mbox {H.c.}}~.
\end{equation}
The supercurrent is therefore a linear combination of terms with 
well defined $H$ and $Q$-charges.

{}In the covariant gauge, we have the
reparametrization ghosts $b$ and $c$, and superconformal ghosts
$\beta$ and $\gamma$ \cite{FMS}. It is most convenient to
bosonize the $\beta,\gamma$
ghosts:
\begin{equation}
\beta = \partial \xi e^{-\phi}, ~~~ \gamma = \eta e^{ \phi}~,
\end{equation}
where $\xi$ and $\eta$ are auxiliary fermions and $\phi$ is a bosonic
ghost field obeying the OPE
$\phi(\overline{z}) \phi(\overline{w}) \sim
{\mbox{log}} ( \overline{z} - \overline{w})$. The conformal dimension of
$e^{q \phi}$
is $-{1\over 2} q (q+2)$.
{}In covariant gauge, vertex operators
are of the form $V(z,{\overline z})=V(z){\overline V}({\overline z})$,
where $V(z)$ and $V(\overline{z})$ are both dimension $1$ operators
constructed from the conformal fields. These include the longitudinal
components as well as the ghosts. The vertex operators for space-time
bosons carry integral ghost charges ($q \in {\bf Z}$) whereas
for space-time fermions the ghost charges are half-integral
($q \in {\bf Z} + {1\over 2}$). Here, $q$ specifies the picture.
The canonical choice is
$q=-1$ for space-time bosons and $q=-{1\over 2}$ for space-time
fermions. We will denote the corresponding vertex operators by
$V_{-1} (z, \overline{z})$ and $V_{-{1\over 2}}(z, \overline{z})$,
respectively. Vertex operators in the $q=0$ picture (with zero ghost charge)
is given by {\em picture changing}~:
\begin{equation}
V_{0}(z,{\overline z})= \lim_{{\overline w}\rightarrow
{\overline z}}{ e^{\phi} T_F ({\overline z})
 V_{-1}(z,{\overline w})}~.
\end{equation}

{}Having constructed the vertex operators for the massless states, one
can in principle compute the scattering amplitudes, or the corresponding
couplings in the superpotential. The coupling of $M$ chiral superfields in
the superpotential is given by the scattering amplitude of the component
fields in the limit when all the external momenta are zero.
Due to holomorphicity, one needs to consider only the scatterings of 
left-handed space-time fermions, with vertices $V_{-1/2}(z,{\overline z})$,
and their space-time superpartners.
Since the total $\phi$ ghost charge in any tree-level correlation 
function is $-2$, it is convenient to choose two of the vertex operators in 
the $-1/2$-picture,
one in the $-1$-picture, and the rest in the $0$-picture.
Using the $SL(2,{\bf C})$ invariance, the scattering amplitude is therefore
\begin{equation}\label{scattem}
 {\cal A}_{M} = g^{M-2}_{\mbox{st}}\int dz_{4} d \overline{z}_{4}
  \cdots dz_{M} d \overline{z}_{M}
        \langle V_{-{1\over 2}}(0,0)V_{-{1\over 2}}(1,1)
          V_{-1}(\infty,\infty) V_{0}(z_4,\overline{z}_{4})
          \cdots V_{0}(z_M,\overline{z}_{M}) \rangle ~,
\end{equation}
where we have normalized the $c$ ghost part of the correlation function
$\langle c(0,0) c(1,1) c(\infty,\infty) \rangle$ to $1$.
For a non-zero coupling, the sum of the $H$-charges as well
as the sum of the $Q$-charges
must be zero in the 
corresponding scattering 
amplitude. Note that the supercurrent carries terms 
with different $H$- and $Q$-charges. Because of picture
changing, $H$- and $Q$- charges are not global charges even though
they must be conserved exactly.  Point group and space 
group selection rules follow from these conservation laws.

\section{${\bf Z}_7$ Orbifold Space Group Selection Rules}\label{space} 

{}In this appendix we derive the space group selection rules for the symmetric 
${\bf Z}_7$ orbifold. 
{\em A priori}, one needs to understand the scattering of the primary 
twist fields and their descendents to solve this problem. 
The general framework for calculating scatterings of twisted fields 
in orbifolds have been
developed in Ref \cite{DFMS} using techniques in conformal field theory. 
The actual calculations, however, can be quite 
non-trivial.
Fortunately, there exists a simpler 
way of deducing the space group selection rules. This other way utilizes 
the techniques recently developed in Ref \cite{KST} based on the bosonic 
supercurrent framework discussed in Appendix \ref{bosonic}. The idea is to 
compute 
the scattering at the enhanced gauge symmetry point where the vertex 
operators for all the twist fields (up to cocycles) can be expressed as 
exponentials of the ${\bf Q}_R$ and ${\bf Q}_L$ charges 
(see Appendix \ref{bosonic}),
hence the problem 
can be solved relatively easily. At the enhanced gauge symmetry point in 
the moduli space the orbifold selection rules are given by the conservation 
of gauge charges ${\bf Q}_L$, and also by the conservation of the 
${\bf Q}_R$ and $H$-charges. (Both of these are affected by picture 
changing, and therefore are neither local nor global charges in space-time.) 
Ultimately, we would like to obtain the orbifold space group selection 
rules {\em away} from the enhanced symmetry point. This can be done by 
considering the corresponding couplings within effective field theory, 
and subsequently breaking the enhanced gauge symmetry by giving vevs to 
the corresponding scalars. The latter procedure is an effective field theory 
manifestation of tuning the stringy moduli away from the enhanced symmetry 
point. Once the enhanced gauge symmetry is broken completely, we obtain the 
space group selection rules ({\em i.e.}, the knowledge of whether a given 
coupling vanishes or not according to this discrete symmetry) by simply 
examining the superpotential. Note that at the enhanced symmetry point 
there are a number of fields in the untwisted sector charged under the 
enhanced gauge symmetry but neutral under the original one. Upon Higgsing 
the enhanced gauge symmetry completely, some of them are eaten in the 
super-Higgs mechanism, and some of them acquire masses via the tree-level 
superpotential. As a result, the number of neutral scalars is 
precisely equal to the dimension of the space parametrized by the geometric 
moduli of the orbifold. This is to be contrasted with the fact that in the 
twisted sectors the number of fields does not depend upon the values of the 
geometric moduli whether they are at a generic or enhanced symmetry point.

{}Thus, let us start from the Narain model with $N=4$ space-time 
supersymmetry in four dimensions. Let the momenta of the
internal (6 right-moving and 22 left-moving) world-sheet bosons span the 
(even self-dual) Narain lattice $\Gamma^{6,22}=\Gamma^{6,6}\otimes\Gamma^{16}$.
Here $\Gamma^{16}$ is the ${\mbox{Spin}}(32)/{\bf Z}_2$ lattice, whereas the 
lattice $\Gamma^{6,6}$ is spanned by the momenta $(p_R \vert\vert p_L)$ with 
$p_L,p_R \in {\tilde \Gamma}^7$ [$SU(7)$ weight lattice], and 
$p_L-p_R \in \Gamma^7$ [$SU(7)$ root lattice].
Note that this corresponds to a compactification on a six-torus with spacial 
values of the constant background metric $g_{ij}$ and (non-zero) 
antisymmetric tensor $B_{ij}$.
This Narain model has gauge group $SU(7) \otimes SO(32)$. The first 
factor $SU(7)$ comes from 
the oscillator excitations and momentum states of the left-moving 
world-sheet bosons corresponding to $\Gamma^{6,6}$ ({\em i.e}, the 
six-torus). The second factor $SO(32)$ comes from the other $16$ 
left-moving world-sheet bosons.

{}Next consider the ${\bf Z}_7$ orbifold model (with non-standard embedding 
of the gauge connection) obtained via twisting the above Narain model by the 
twist $T_7$ given in section III.
This model has $N=1$ supersymmetry, and gauge group 
$U(1)^6 \otimes [U(4)^3\otimes SO(8)]$. The factor $U(4)^3\otimes SO(8)$ 
comes from the breaking of $SO(32)$. The factor $U(1)^6$ comes from the 
breaking of $SU(7)$. Note that, as discussed in Appendix \ref{bosonic}, we can 
represent the ${\bf Z}_7$ twist 
$(\theta,\theta^2,\theta^3 \vert\vert \theta,\theta^2,\theta^3)$ acting 
in the six-torus in terms of a ${\bf Z}_7$ ({\em i.e.}, order $7$) shift 
provided that the right-moving supercurrent is written in the bosonized form. 
Here we give this shift in terms of the $SU(7)\supset U(1)^6$ basis, where 
the normalization radii of the six $U(1)$'s are given by
$({1\over{\sqrt{1\cdot 2}}}, {1\over{\sqrt{2\cdot 3}}}, 
{1\over{\sqrt{3\cdot 4}}},  {1\over{\sqrt{4\cdot 5}}}, 
{1\over{\sqrt{5\cdot 6}}}, {1\over{\sqrt{6\cdot 7}}})$. Thus, in this basis 
the twist $T_7$ is replaced by the shift:
\begin{equation}
 T^\prime_7=(-\textstyle{1\over 7},-{3\over 7}-{6\over 7},-{10\over 7},
 -{15\over 7},
 -{21\over 7}\vert\vert
 -{1\over 7},-{3\over 7}-{6\over 7},-{10\over 7},-{15\over 7},-{21\over 7}
  \vert 
 ({1\over 7})^4 ({2\over 7})^4 ({3\over 7})^4 0^4)~.
\end{equation}
In this basis it is straightforward to work out the ${\bf Q}_R$ and 
${\bf Q}_L$ charges of the massless states of the model. The latter are 
the same as in the model discussed in Sec. III, except for the untwisted 
sector singlets. Thus, instead of three neutral singlets $\phi_a$ 
(see Table II) we have $21$ fields $\phi^a_\alpha$, $\alpha=1,...,7$, that 
are singlets under $U(4)^3 \otimes SO(8)$ gauge group, but are charged under 
$U(1)^6$ Abelian subgroup. Their charges are given in Table III. By giving 
vevs to these singlets we can completely break $U(1)^6$ gauge symmetry. 
Due to the super-Higgs mechanism and the corresponding superpotential after 
Higgsing, only three neutral fields $\phi_a$ survive in the massless spectrum. 
The rest are either eaten by the gauge bosons or become heavy via the 
couplings in the superpotential. This field theory breaking is in one-to-one 
correspondence with the string theory picture of moving in the moduli space 
$[SU(1,1,{\bf Z})\backslash SU(1,1)/U(1)]^3$ discussed earlier. That is, 
we are moving the moduli away from the special point of enhanced gauge 
symmetry into the bulk, {\em i.e.}, to some generic point.

{}The bosonic supercurrent is given by 
\begin{equation}
T_F = {1\over 2 \sqrt{7}} \left(
       e^{i \rho_1} \sum_{\alpha=1}^{7} e^{iQ^1_\alpha \cdot \phi}
     + e^{i \rho_2} \sum_{\alpha=1}^{7} e^{iQ^2_\alpha \cdot \phi}
     + e^{i \rho_3} \sum_{\alpha=1}^{7} e^{-iQ^3_\alpha \cdot \phi}\right)
     + {\mbox{H.c.}}~,
\end{equation}
where the $Q^a_\alpha$ charges 
for the currents $i\partial X^a$ are the same as the ${\bf Q}_L$ charges 
for the fields $\phi^a_\alpha$ in Table III (and this is no coincidence for 
the orbifold is symmetric).

{}Note that the untwisted sector fields $P_a$, $Q_a$, $R_a$ and $\Phi_a$ are 
not charged under the enhanced $U(1)^6$ gauge symmetry, so that the couplings 
$\lambda_{1,2,3}$ for the untwisted sector fields do not vanish at any point 
in the moduli space (but smoothly vary with the moduli). The twisted sector 
fields $T^a_\alpha$ and $S^a_\alpha$ do carry $U(1)^6$ charges. 
The $U(1)^6$ charges along 
with the ${\bf Q}_R$ charges for the
fields $T^a_\alpha$ and $S^a_\alpha$
are given in Tables IV and V. 
Because $T^a_\alpha$ and 
$S^a_\alpha$ carry $U(1)^6$ charges, some of the couplings 
$\Lambda^{\alpha\beta\gamma}$ that are non-zero at generic points vanish 
at the enhanced gauge symmetry point. 

{}Since the model possesses explicit ${\bf Z}_3$ cyclic symmetry 
$(a=1)\rightarrow (a=2)\rightarrow (a=3)\rightarrow (a=1)$, we can confine 
our attention to couplings $(Q_1)^2
T^1_\alpha T^2_\beta S^3_\gamma$. For example, according to Tables III, 
IV and V, the coupling $(Q_1)^2 T^1_\alpha T^2_\beta S^3_\gamma$ 
for $\alpha=1$, $\beta= \gamma=4$ is allowed by ${\bf Q}_R$ (here one needs 
to take into account the picture changing) and ${\bf Q}_L$ charge 
conservation. On the other hand, say, the coupling with 
$\alpha=\beta= \gamma=1$ is not allowed. There is, however, a higher point 
coupling, namely, 
$(Q_1)^2 T^1_1 T^2_1 S^3_1 \phi^1_3 \phi^3_2$ that is allowed. Upon the 
fields $\phi^1_3$ and $\phi^3_2$ acquiring vevs, we, therefore, have an 
effective coupling $(Q_1)^2 T^1_1 T^2_1 S^3_1$. From examining the 
${\bf Q}_R$ and ${\bf Q}_L$ charge conservation in the scattering of 
states $(Q_1)^2 T^1_\alpha T^2_\beta S^3_\gamma$, it becomes clear that 
near the enhanced symmetry point upon the fields $S^3_\alpha$ and $Q_1$ 
acquiring vevs, all the fields $T^1_\alpha$ and $T^2_\beta$ generically 
become massive. Similarly, if all the vevs $Q_a$ and $S^a_\alpha$ are 
non-zero, all the fields $T^a_\alpha$ are generically massive.
In fact, this conclusion does not depend on being close to the enhanced 
symmetry point. Thus, consider the basis for $\alpha,\beta,\gamma$ indices 
such that they label the fixed points of the orbifold [this basis is 
{\em not} the same as that of $SU(7)\supset U(1)^6$, but can be constructed 
from the latter via a rotation]. Then it is clear that in the limit of large 
volume of the orbifold the couplings $\Lambda^{\alpha\beta\gamma}$ for 
$\alpha,\beta,\gamma$ are exponentially suppressed, whereas the couplings 
$\Lambda^{\alpha\alpha\alpha}$
are not. The latter couplings are non-zero at generic points in the moduli
space. From this it should become clear that generically all the fields 
$T^a_\alpha$ are heavy as long as we turn on vevs for all of the fields 
$Q_a$, and also $S^a_\alpha$.

\begin{table}[t]
\begin{tabular}{|c|c|l|l|}
Sector & Field & $SU(4)\otimes SU(4)\otimes SU(4)\otimes SO(8)\otimes U(1)^3$ 
       & Comments \\
\hline
Closed & & &\\
Untwisted & $\phi_{a}$ & $3({\bf 1}, {\bf 1}, {\bf 1}, {\bf 1})(0,0,0)_L$ 
          & $a=1,2,3$\\
\hline
Closed & $S_{\alpha}^a$ & $21({\bf 1}, {\bf 1}, {\bf 1}, {\bf 1})(0,0,0)_L$ 
         & $a=1$ to $3$ \\
Twisted & & & $\alpha=1$ to $7$ \\
\hline
    & $P_1$ & $({\bf 4}, {\bf 1}, {\bf 1}, {\bf 8}_v)(+1,0,0)_L$ & \\
    & $P_2$ & $({\bf 1}, {\bf 4}, {\bf 1}, {\bf 8}_v)(0,+1,0)_L$ & \\
    & $P_3$ & $({\bf 1}, {\bf 1}, \overline{\bf 4}, {\bf 8}_v)(0,0,-1)_L$ & \\
&$Q_1$&$({\bf 1}, \overline{\bf 4}, {\bf 4}, {\bf 1})(0,-1,+1)_L$ & \\
&$Q_2$&$(\overline{\bf 4}, {\bf 1}, {\bf 4}, {\bf 1})(-1,0,+1)_L$ & \\
Open
&$Q_3$&$(\overline{\bf 4}, \overline{\bf 4}, {\bf 1}, {\bf 1})(-1,-1,0)_L$ &\\
   & $R_1$ & $({\overline{\bf 4}}, {\bf 4}, {\bf 1}, {\bf 1})(-1,+1,0)_L$ & \\
   & $R_2$ & $({\bf 1}, {\overline{\bf 4}}, \overline{\bf 4}, 
               {\bf 1})(0,-1,-1)_L$  & \\
   & $R_3$ & $({\bf 4}, {\bf 1}, {\bf 4}, {\bf 1})(+1,0,+1)_L$ & \\
& $\Phi_1$ & $({\bf 1}, {\bf 1}, {\bf 6}, {\bf 1})(0,0,-2)_L$ & \\
& $\Phi_2$ & $({\bf 6}, {\bf 1}, {\bf 1}, {\bf 1})(+2,0,0)_L$ & \\
& $\Phi_3$ & $({\bf 1}, {\bf 6}, {\bf 1}, {\bf 1})(0,+2,0)_L$ &\\
\end{tabular}
\caption{The massless spectrum of the type I model with $N=1$ space-time 
supersymmetry and gauge group 
$SU(4)\otimes SU(4) \otimes SU(4) \otimes SO(8) \otimes U(1)^3$ discussed in 
section II. The gravity, dilaton and gauge supermultiplets are not shown.}  
\end{table}

\begin{table}[t]
\begin{tabular}{|c|c|l|l|l|}
Sector & Field 
& $SU(4)^{3} \otimes SO(8) \otimes U(1)^3$ 
& $(H_1,H_2,H_3)_{-1}$ & $(H_1,H_2,H_3)_{-1/2}$ \\
\hline
  & $\phi_{1}$ & $({\bf 1}, {\bf 1}, {\bf 1}, {\bf 1})(0,0,0)_L$ 
  & $(-1,0,0)$  & $(-{1\over 2},+{1\over 2},-{1\over 2})$ \\
  & $\phi_{2}$ & $({\bf 1}, {\bf 1}, {\bf 1}, {\bf 1})(0,0,0)_L$ 
  & $(0,-1,0)$  & $(+{1\over 2},-{1\over 2},-{1\over 2})$ \\
  & $\phi_{3}$ & $({\bf 1}, {\bf 1}, {\bf 1}, {\bf 1})(0,0,0)_L$ 
  & $(0,0,+1)$  & $(+{1\over 2},+{1\over 2},+{1\over 2})$   \\
    & $P_1$ & $({\bf 4}, {\bf 1}, {\bf 1}, {\bf 8}_v)(+1,0,0)_L$ 
    & $(-1,0,0)$  & $(-{1\over 2},+{1\over 2},-{1\over 2})$ \\
    & $P_2$ & $({\bf 1}, {\bf 4}, {\bf 1}, {\bf 8}_v)(0,+1,0)_L$  
    & $(0,-1,0)$  & $(+{1\over 2},-{1\over 2},-{1\over 2})$ \\
    & $P_3$ & $({\bf 1}, {\bf 1}, \overline{\bf 4}, {\bf 8}_v)(0,0,-1)_L$  
    & $(0,0,+1)$  & $(+{1\over 2},+{1\over 2},+{1\over 2})$   \\
&$Q_1$&$({\bf 1}, \overline{\bf 4}, {\bf 4}, {\bf 1})(0,-1,+1)_L$
      & $(-1,0,0)$  & $(-{1\over 2},+{1\over 2},-{1\over 2})$ \\
Untwisted
&$Q_2$&$(\overline{\bf 4}, {\bf 1}, {\bf 4}, {\bf 1})(-1,0,+1)_L$
& $(0,-1,0)$  & $(+{1\over 2},-{1\over 2},-{1\over 2})$ \\
&$Q_3$&$(\overline{\bf 4}, \overline{\bf 4}, {\bf 1}, {\bf 1})(-1,-1,0)_L$
& $(0,0,+1)$  & $(+{1\over 2},+{1\over 2},+{1\over 2})$   \\
   & $R_1$ & $({\overline{\bf 4}}, {\bf 4}, {\bf 1}, {\bf 1})(-1,+1,0)_L$ 
   & $(-1,0,0)$  & $(-{1\over 2},+{1\over 2},-{1\over 2})$ \\
   & $R_2$ & $({\bf 1}, {\overline{\bf 4}}, \overline{\bf 4}, 
               {\bf 1})(0,-1,-1)_L$  
   & $(0,-1,0)$  & $(+{1\over 2},-{1\over 2},-{1\over 2})$ \\
   & $R_3$ & $({\bf 4}, {\bf 1}, {\bf 4}, {\bf 1})(+1,0,+1)_L$  
   & $(0,0,+1)$  & $(+{1\over 2},+{1\over 2},+{1\over 2})$   \\
& $\Phi_1$ & $({\bf 1}, {\bf 1}, {\bf 6}, {\bf 1})(0,0,-2)_L$
& $(-1,0,0)$  & $(-{1\over 2},+{1\over 2},-{1\over 2})$ \\
& $\Phi_2$ & $({\bf 6}, {\bf 1}, {\bf 1}, {\bf 1})(+2,0,0)_L$ 
& $(0,-1,0)$  & $(+{1\over 2},-{1\over 2},-{1\over 2})$ \\
& $\Phi_3$ & $({\bf 1}, {\bf 6}, {\bf 1}, {\bf 1})(0,+2,0)_L$
 & $(0,0,+1)$  & $(+{1\over 2},+{1\over 2},+{1\over 2})$   \\
\hline
Twisted & $S_{\alpha}^1$ 
      & $7({\bf 1}, {\bf 1}, {\bf 1}, {\bf 1})(4/7, 8/7, 12/7)_L$ 
& $(-{1\over 7},-{2\over 7},+{4\over 7})$ 
& $(+{5\over 14},+{3\over 14},+{1\over 14})$ \\
$\theta$, $\theta^6$
   & $T_{\alpha}^1$ 
   & $7({\bf 1}, {\bf 1}, {\bf 6}, {\bf 1})(4/7, 8/7, -2/7)_L$ 
& $(-{1\over 7},-{2\over 7},+{4\over 7})$ 
& $(+{5\over 14},+{3\over 14},+{1\over 14})$ \\
\hline
Twisted & $S_{\alpha}^2$ 
  & $7({\bf 1}, {\bf 1}, {\bf 1}, {\bf 1})({8/7}, -{12/7}, -{4/7})_L$ 
& $(-{2\over 7},-{4\over 7},+{1\over 7})$ 
& $(+{3\over 14},-{1\over 14},-{5\over 14})$ \\
$\theta^2$, $\theta^5$
   & $T_{\alpha}^2$ 
   & $7({\bf 1}, {\bf 6}, {\bf 1}, {\bf 1})(8/7, 2/7, -4/7)_L$ 
& $(-{2\over 7},-{4\over 7},+{1\over 7})$ 
& $(+{3\over 14},-{1\over 14},-{5\over 14})$ \\
\hline
Twisted & $S_{\alpha}^3$ 
& $7({\bf 1}, {\bf 1}, {\bf 1}, {\bf 1})(-{12/7}, {4/7}, -{8/7})_L$ 
& $(-{4 \over 7},-{1\over 7},+{2 \over 7})$ 
& $(-{1\over 14},+{5\over 14},-{3\over 14})$ \\
$\theta^3$, $\theta^4$
   & $T_{\alpha}^3$ 
   & $7({\bf 6}, {\bf 1}, {\bf 1}, {\bf 1})({2/7}, {4/7}, -{8/7})_L$ 
& $(-{4 \over 7},-{1 \over 7},+{2 \over 7})$ 
& $(-{1\over 14},+{5\over 14},-{3\over 14})$ \\
\end{tabular}
\caption{The massless spectrum of the heterotic model with $N=1$ space-time 
supersymmetry and gauge group 
$SU(4)\otimes SU(4) \otimes SU(4) \otimes SO(8) \otimes U(1)^3$ 
discussed in section III. The $H$-charges in both the $-1$ picture and
the $-1/2$ picture are also given. The gravity, dilaton and gauge 
supermultiplets are 
not shown.}  
\end{table}

\begin{table}[t]
\begin{tabular}{|c|l|l|}
Field & ${\bf Q}_R$ & ${\bf Q}_L$ \\
\hline
 $\phi^1_{\alpha}$ & $(0,0,0,0,0,0)$ & $(+1,+1,+1,+1,+1,+7)$ \\
                                  & $(0,0,0,0,0,0)$ & $(-2,0,0,0,0,0)$ \\
                                  & $(0,0,0,0,0,0)$ & $(+1,-3,0,0,0,0)$ \\
                                  & $(0,0,0,0,0,0)$ & $(0,+2,-4,0,0,0)$ \\
                                  & $(0,0,0,0,0,0)$ & $(0,0,+3,-5,0,0)$ \\
                                  & $(0,0,0,0,0,0)$ & $(0,0,0,+4,-6,0)$ \\
                                  & $(0,0,0,0,0,0)$ & $(0,0,0,0,+5,-7)$ \\
\hline
 $\phi^2_{\alpha}$ & $(0,0,0,0,0,0)$ & $(+1,+1,+1,+1,+6,0)$ \\
                                  & $(0,0,0,0,0,0)$ & $(-1,+1,+1,+1,+1,+7)$ \\
                                  & $(0,0,0,0,0,0)$ & $(-1,-3,0,0,0,0)$ \\
                                  & $(0,0,0,0,0,0)$ & $(+1,-1,-4,0,0,0)$ \\
                                  & $(0,0,0,0,0,0)$ & $(0,+2,-1,-5,0,0)$ \\
                                  & $(0,0,0,0,0,0)$ & $(0,0,+3-1,-6,0)$ \\
                                  & $(0,0,0,0,0,0)$ & $(0,0,0,+4,-1,-7)$ \\
\hline
 $\phi^3_{\alpha}$ & $(0,0,0,0,0,0)$ & $(+1,+1,+4,0,0,0)$ \\
                                  & $(0,0,0,0,0,0)$ & $(-1,+1,+1,+5,0,0)$ \\
                                  & $(0,0,0,0,0,0)$ & $(0,-2,+1,+1,+6,0)$ \\
                                  & $(0,0,0,0,0,0)$ & $(0,0,-3,+1,+1,+7)$ \\
                                  & $(0,0,0,0,0,0)$ & $(-1,-1,-1,-5,0,0)$ \\
                                  & $(0,0,0,0,0,0)$ & $(+1,-1,-1-1,-6,0)$ \\
                                  & $(0,0,0,0,0,0)$ & $(0,+2,-1,-1,-1,-7)$ \\
\hline
 & $({1\over{\sqrt{1\cdot 2}}}, {1\over{\sqrt{2\cdot 3}}}, 
    {1\over{\sqrt{3\cdot 4}}},
 {1\over{\sqrt{4\cdot 5}}}, {1\over{\sqrt{5\cdot 6}}}, 
 {1\over{\sqrt{6\cdot 7}}})$ & $({1\over{\sqrt{1\cdot 2}}}, 
 {1\over{\sqrt{2\cdot 3}}}, {1\over{\sqrt{3\cdot 4}}},
 {1\over{\sqrt{4\cdot 5}}}, {1\over{\sqrt{5\cdot 6}}}, 
 {1\over{\sqrt{6\cdot 7}}})$ \\
\end{tabular}
\caption{The ${\bf Q}_R$ and ${\bf Q}_L$ charges for the untwisted sector 
fields $\phi^a_\alpha$. The $U(1)^6_R$ and $U(1)^6_L$ normalization radii 
are given at the bottom of the Table.}  
\end{table}

\begin{table}[t]
\begin{tabular}{|c|l|l|}
Field & ${\bf Q}_R$ & ${\bf Q}_L$ \\
\hline
 $T^1_{\alpha}$     &  ${1\over 7}(-1,-3,-6,-10,-15,-21)$& 
                       ${1\over 7}(-1,-3,-6,-10,-15,-21)$ \\
                    & ${1\over 7}(+6,+4,+1,-3,-8,-14)$&  
                      ${1\over 7}(+6,+4,+1,-3,-8,-14)$ \\
                    & ${1\over 7}(-1,-3,-6,-10,-15,+21)$& 
                      ${1\over 7}(-1,-3,-6,-10,-15,+21)$\\
                    & ${1\over 7}(-1,+11,+8,+4,-1,-7)$&  
                      ${1\over 7}(-1,+11,+8,+4,-1,-7)$\\
                    & ${1\over 7}(-1,-3,-6,-10,+20,+14)$ &
                      ${1\over 7}(-1,-3,-6,-10,+20,+14)$\\
                    & ${1\over 7}(-1,-3,+15,+11,+6,0)$&  
                      ${1\over 7}(-1,-3,+15,+11,+6,0)$\\
                    & ${1\over 7}(-1,-3,-6,+18,+13,+7)$ & 
                      ${1\over 7}(-1,-3,-6,+18,+13,+7)$\\
\hline
 $T^2_{\alpha}$     &  ${1\over 7}(-2,+8,+2,-6,+19,+7)$& 
                       ${1\over 7}(-2,+8,+2,-6,+19,+7)$\\
                    & ${1\over 7}(-2,+8,+2,-6,-16,+14)$&
                      ${1\over 7}(-2,+8,+2,-6,-16,+14)$ \\
                    & ${1\over 7}(+5,+1,-5,-13,+12,0)$&
                      ${1\over 7}(+5,+1,-5,-13,+12,0)$\\
                    & ${1\over 7}(-2,-6,+9,+1,-9,+21)$& 
                      ${1\over 7}(-2,-6,+9,+1,-9,+21)$\\
                    & ${1\over 7}(+5,+1,-5,+15,+5,-7)$ &
                      ${1\over 7}(+5,+1,-5,+15,+5,-7)$\\
                    & ${1\over 7}(-2,-6,+9,+1,-9,-21)$& 
                      ${1\over 7}(-2,-6,+9,+1,-9,-21)$\\
                    & ${1\over 7}(-2,-6,-12,+8,-2,-14)$ &
                      ${1\over 7}(-2,-6,-12,+8,-2,-14)$ \\
\hline 
 $T^3_{\alpha}$     &  ${1\over 7}(+3,-5,+4,+16,-4,+14)$& 
                       ${1\over 7}(+3,-5,+4,+16,-4,+14)$\\
                    & ${1\over 7}(+3,+9,-3,+9,-11,+7)$& 
                      ${1\over 7}(+3,+9,-3,+9,-11,+7)$\\
                    & ${1\over 7}(+3,-5,+4,-12,+3,+21)$&
                      ${1\over 7}(+3,-5,+4,-12,+3,+21)$\\
                    & ${1\over 7}(-4,+2,-10,+2,-18,0)$& 
                      ${1\over 7}(-4,+2,-10,+2,-18,0)$\\
                    & ${1\over 7}(+3,-5,+4,-12,+3,-21)$ &
                      ${1\over 7}(+3,-5,+4,-12,+3,-21)$\\
                    & ${1\over 7}(-4,+2,-10,+2,+17,-7)$& 
                      ${1\over 7}(-4,+2,-10,+2,+17,-7)$\\
                    & ${1\over 7}(-4,+2,+11,-5,+10,-14)$ &
                      ${1\over 7}(-4,+2,+11,-5,+10,-14)$ \\
\hline 
 & $({1\over{\sqrt{1\cdot 2}}}, {1\over{\sqrt{2\cdot 3}}}, 
     {1\over{\sqrt{3\cdot 4}}},
 {1\over{\sqrt{4\cdot 5}}}, {1\over{\sqrt{5\cdot 6}}}, 
 {1\over{\sqrt{6\cdot 7}}})$ & $({1\over{\sqrt{1\cdot 2}}}, 
 {1\over{\sqrt{2\cdot 3}}}, {1\over{\sqrt{3\cdot 4}}},
 {1\over{\sqrt{4\cdot 5}}}, {1\over{\sqrt{5\cdot 6}}}, 
 {1\over{\sqrt{6\cdot 7}}})$ \\
\end{tabular}
\caption{The ${\bf Q}_R$ and ${\bf Q}_L$ charges for the twisted sector 
fields $T^a_\alpha$ . The $U(1)^6_R$ and $U(1)^6_L$ normalization radii 
are given at the bottom of the Table.}  
\end{table}

\begin{table}[t]
\begin{tabular}{|c|l|l|}
Field & ${\bf Q}_R$ & ${\bf Q}_L$ \\
\hline
 $S^1_{\alpha}$     &  ${1\over 7}(-1,-3,-6,-10,-15,-21)$& 
                       ${1\over 7}(+6,+4,+1,-3,-8,+28)$ \\
                    & ${1\over 7}(+6,+4,+1,-3,-8,-14)$& 
                      ${1\over 7}(-8,+4,+1,-3,-8,-14)$ \\
                    & ${1\over 7}(-1,-3,-6,-10,-15,+21)$&
                      ${1\over 7}(-1,-3,-6,-10,+20,-28)$\\
                    & ${1\over 7}(-1,+11,+8,+4,-1,-7)$& 
                      ${1\over 7}(+6,-10,+8,+4,-1,-7)$\\
                    & ${1\over 7}(-1,-3,-6,-10,+20,+14)$ &
                      ${1\over 7}(-1,-3,-6,+18,-22,+14)$\\
                    & ${1\over 7}(-1,-3,+15,+11,+6,0)$& 
                      ${1\over 7}(-1,+11,-13,+11,+6,0)$\\
                    & ${1\over 7}(-1,-3,-6,+18,+13,+7)$ & 
                      ${1\over 7}(-1,-3,+15,-17,+13,+7)$\\
\hline
 $S^2_{\alpha}$     &  ${1\over 7}(-2,+8,+2,-6,+19,+7)$& 
                       ${1\over 7}(+5,+1,-5,-13,-23,+7)$\\
                    & ${1\over 7}(-2,+8,+2,-6,-16,+14)$&
                      ${1\over 7}(-2,-6,+9,+1,+26,+14)$ \\
                    & ${1\over 7}(+5,+1,-5,-13,+12,0)$&
                      ${1\over 7}(-2,+8,+2,+22,+12,0)$\\
                    & ${1\over 7}(-2,-6,+9,+1,-9,+21)$& 
                      ${1\over 7}(-2,+8,+2,-6,-16,-28)$\\
                    & ${1\over 7}(+5,+1,-5,+15,+5,-7)$ &
                      ${1\over 7}(-2,-6,-12,-20,+5,-7)$\\
                    & ${1\over 7}(-2,-6,+9,+1,-9,-21)$& 
                      ${1\over 7}(-2,-6,-12,+8,-2,+28)$\\
                    & ${1\over 7}(-2,-6,-12,+8,-2,-14)$ &
                      ${1\over 7}(+5,+1,+16,+8,-2,-14)$ \\
\hline 
 $S^3_{\alpha}$     &  ${1\over 7}(+3,-5,+4,+16,-4,+14)$& 
                       ${1\over 7}(+3,+9,-3,-19,-4,+14)$\\
                    & ${1\over 7}(+3,+9,-3,+9,-11,+7)$& 
                      ${1\over 7}(-4,-12,-3,+9,-11,+7)$\\
                    & ${1\over 7}(+3,-5,+4,-12,+3,+21)$&
                      ${1\over 7}(+3,-5,+4,+16,-4,-28)$\\
                    & ${1\over 7}(-4,+2,-10,+2,-18,0)$& 
                      ${1\over 7}(+3,+9,-3,+9,+24,0)$\\
                    & ${1\over 7}(+3,-5,+4,-12,+3,-21)$ &
                      ${1\over 7}(-4,+2,+11,-5,+10,+28)$\\
                    & ${1\over 7}(-4,+2,-10,+2,+17,-7)$& 
                      ${1\over 7}(-4,+2,+11,-5,-25,-7)$\\
                    & ${1\over 7}(-4,+2,+11,-5,+10,-14)$ &
                      ${1\over 7}(+3,-5,-17,-5,+10,-14)$ \\
\hline 
 & $({1\over{\sqrt{1\cdot 2}}}, {1\over{\sqrt{2\cdot 3}}}, 
    {1\over{\sqrt{3\cdot 4}}},
 {1\over{\sqrt{4\cdot 5}}}, {1\over{\sqrt{5\cdot 6}}}, 
 {1\over{\sqrt{6\cdot 7}}})$ & $({1\over{\sqrt{1\cdot 2}}}, 
 {1\over{\sqrt{2\cdot 3}}}, {1\over{\sqrt{3\cdot 4}}},
 {1\over{\sqrt{4\cdot 5}}}, {1\over{\sqrt{5\cdot 6}}}, 
 {1\over{\sqrt{6\cdot 7}}})$ \\
\end{tabular}
\caption{The ${\bf Q}_R$ and ${\bf Q}_L$ charges for the twisted sector 
fields and $S^a_\alpha$. The $U(1)^6_R$ and $U(1)^6_L$ normalization radii 
are given at the bottom of the Table.}  
\end{table}

\newpage
\begin{figure}[t]
\epsfxsize=16 cm
\epsfbox{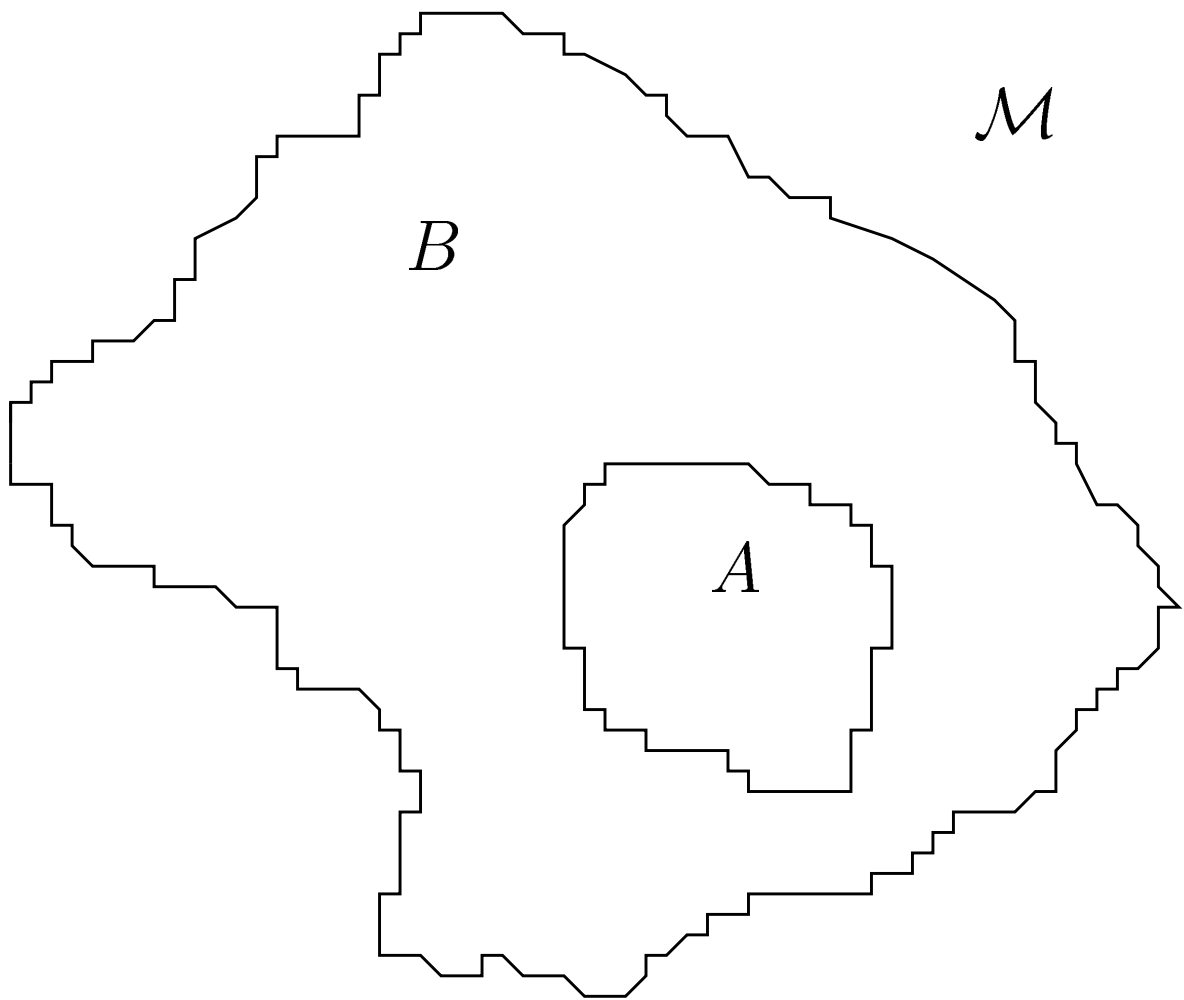}
\caption{A schematic picture of the (perturbative) moduli space ${\cal M}$
(of the heterotic model). Region $A$ is the subspace corresponding to the 
type I model. Region $B$ (that complements $A$ in ${\cal M}$) is the subspace 
where some or all of the $S^a_{\alpha}$ vevs are zero and some or all of the 
$T^a_{\alpha}$ fields are massless.}
\end{figure}

\end{document}